\documentclass[12pt]{article}
\usepackage[dvips]{graphicx}
\usepackage{fancyhdr}
\newcommand{\prl}[2]{{\em Phys. Rev. Lett.}         {\bf  #1}, #2 }
\newcommand{\prd}[2]{{\em Phys. Rev.}               {\bf D#1}, #2 }
\newcommand{\sci}[2]{{\em Science}                  {\bf  #1}, #2 }
\newcommand{\pr }[2]{{\em Phys. Rep.}               {\bf  #1}, #2 }
\newcommand{\etal}{{\em et al.}}

\newcommand{\url}[1]{{\tt #1}}

\begin{document}

\begin{centering}
{\Large \bf Qweak: A Precision Measurement of the \\ 
                   Proton's Weak Charge
}

{\large Gregory S.\ Mitchell, for the Qweak Collaboration$^\dagger$}

{\em
            Physics Division, 
            Los Alamos National Laboratory, 
            Los Alamos, NM 87545, USA
}

\end{centering}

\begin{abstract}
\small
The Qweak experiment at Jefferson Lab aims to make a 4\%
measurement of the parity-violating asymmetry in elastic
scattering at very low $Q^2$ of a longitudinally polarized 
electron beam on a proton target.  The experiment will measure 
the weak charge of the proton, and thus the weak mixing angle at 
low energy scale, providing a precision test of the Standard Model.  
Since the value of the weak mixing angle is approximately 1/4, the
weak charge of the proton $Q_w^p = 1-4 \sin^2 \theta_w$ is 
suppressed in the Standard Model, making it especially sensitive 
to the value of the mixing angle and also to possible new physics.
The experiment is approved to run at JLab, and the construction
plan calls for the hardware to be ready to install in Hall C in 2007.  
The experiment will be a 2200 hour measurement, employing: 
an 80\% polarized,  180 $\mu$A, 1.2 GeV electron beam; 
a 35 cm liquid hydrogen target;
and a toroidal magnet to focus electrons scattered at 9$^\circ$, 
a small forward angle corresponding to $Q^2$ = 0.03 (GeV/c)$^2$. 
With these kinematics the systematic
uncertainties from hadronic processes are strongly suppressed.
To obtain the necessary statistics the experiment
must run at an event rate of over 6 GHz.
This requires current mode detection of the scattered electrons,
which will be achieved with synthetic quartz {\u C}erenkov detectors. 
A tracking system will be used in a low-rate counting mode to 
determine average $Q^2$ and the dilution factor of background events.  
The theoretical context of the experiment and the status of 
its design are discussed.

\end{abstract}

%%%%%%%%%%%%%%%%%%%%%%%%%%%%%%%%%%%%%%%%%%%%
%% MAINMATTER
%%%%%%%%%%%%%%%%%%%%%%%%%%%%%%%%%%%%%%%%%%%%

\centerline{\bf INTRODUCTION}

%The Standard Model provides a largely successful picture 
%of elementary particles and their interactions.  
There are strong theoretical 
reasons to expect that the Standard Model is a low-energy effective
theory of some more fundamental description of nature.
To identify new physics, one method is to observe directly
new particles and interactions at large energy scale.  
Alternatively, an indirect search can be made at low energy, 
where a precision measurement may observe small effects due to
the new physics.

The Qweak experiment \cite{Qweak} at Jefferson Lab (JLab)
will make such a precision measurement of the
asymmetry between cross-sections for positive and negative helicity 
electrons in polarized elastic electron-proton scattering.
The asymmetry violates parity, and arises 
from the interference of electromagnetic and weak amplitudes 
(photon and $Z$ boson exchange).
At this low energy scale, the asymmetry is a measure of the
weak charge of the proton, ${\rm Q}^p_w$, which is
the strength of the weak vector coupling of the $Z$ boson
to the proton.
In the limit of small scattering angle and small momentum transfer
($Q^2 \rightarrow 0$), the asymmetry is given by~\cite{Musolf}:
$$\frac{\sigma_+ - \sigma_-}{\sigma_+ + \sigma_-} =
[ \frac{-G_F}{4 \pi \alpha \sqrt{2}}  ]
[ Q^2 {\rm Q}^p_w + Q^4 B(Q^2) ] \ 
\approx \ -0.3 \ {\rm ppm}\ {\rm at}\  Q^2 = 0.03\  {\rm GeV}^2 $$
where $B(Q^2)$ is a contribution from electromagnetic and weak
form factors.  
To lowest order, the weak charge is
${\rm Q}^p_w = 1 - 4 \sin^2 \theta_w$, where $\sin^2  \theta_w
\approx 0.23$ is the weak mixing angle.  
The goal of the Qweak experiment is a 4\% measurement of
${\rm Q}^p_w$, which corresponds to a 0.3\% measurement of 
$\sin^2  \theta_w$. 

\begin{figure}
\includegraphics[height=.65\textwidth]{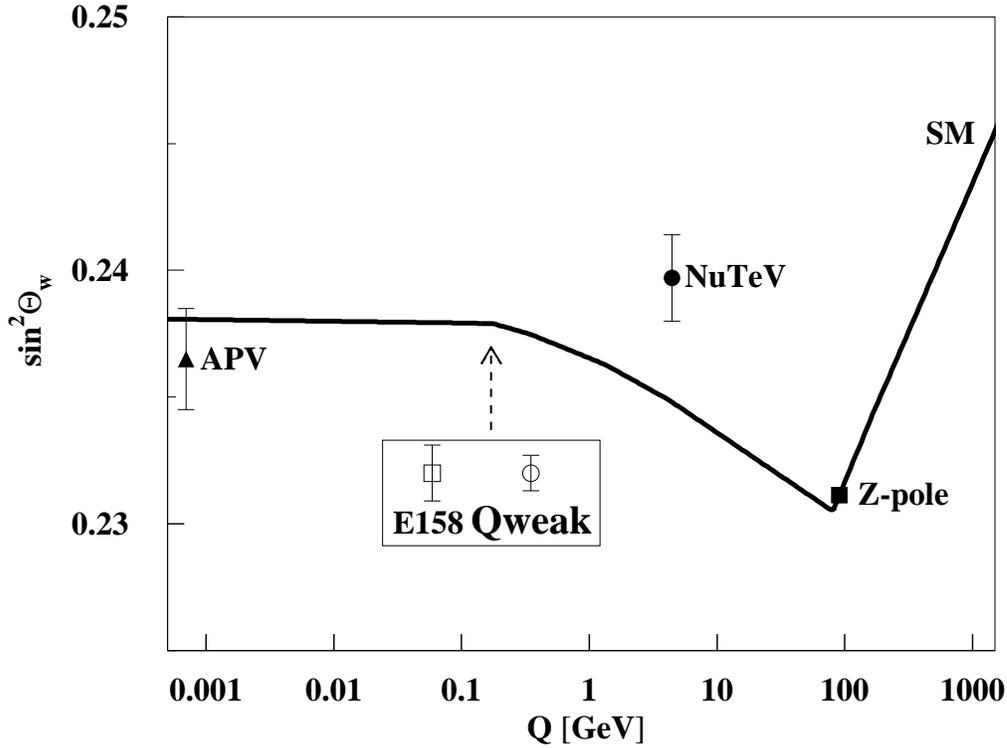}
\caption{Running of the weak mixing angle in the Standard Model,
calculated in the $\overline{\mathrm{MS}}$ scheme.  
Shown are results from atomic parity
violation~\cite{Edwards,Vetter,Wood}, 
NuTeV~\cite{Zeller}, 
and the $Z$ pole~\cite{PDG}.  
E158~\cite{E158} is a currently running 
experiment at SLAC, and the error bar shown is the E158 proposal goal.  
Figure courtesy of J.\ Erler,
A.\ Kurylov, and M.J.\ Ramsey-Musolf~\cite{Erler}.}
\label{fig:sin2theta}
\end{figure}

The weak mixing angle is the single most important parameter of the
Standard Model.
The Qweak experiment will be a low-energy measurement of the weak 
mixing angle, which is a test of the Standard Model running of 
the angle~\cite{Erler, Kurylov}.  As shown in Fig.~\ref{fig:sin2theta}
the value of the weak mixing angle is predicted to 
change (in the $\overline{\mathrm{MS}}$ renormalization scheme) by
$\sim4\%$ from the energy scale of the $Z$ pole (where a decade 
of precision measurements has been made at high
energy colliders at SLAC and CERN~\cite{PDG}) to the low energy scale of
Qweak.  With a 0.3\% measurement of the weak mixing angle,
the Qweak experiment will make a $\sim10\sigma$ verification of this 
effect.

\rfoot{\thepage} %\pagestyle{empty} 
% this command on page 2, so the 'submitted to' comment
% is only on the first page.

Weak charge and mixing angle results can also be obtained 
in other types of experiments.  
In atomic parity violation experiments~\cite{Edwards,Vetter,Wood}, 
weak charge values are complicated to extract 
due to atomic structure and many-body nuclear effects
(see, for example~\cite{Kuchiev,Milstein}).
The NuTeV experiment~\cite{Zeller} at Fermilab has seen evidence of 
a non-Standard Model value for the weak mixing angle at a 
lower energy scale than the $Z$ pole, but its result is sensitive 
to choice of parton distributions and radiative corrections.
A currently running experiment at SLAC, E158~\cite{E158}, 
is measuring parity violation in electron-electron scattering,
and is complementary to Qweak in its sensitivity to different types of new
physics (for example: SUSY, leptoquarks, $Z'$ boson, R-parity
violation, \ldots) \cite{Erler,Kurylov}.
The scientific impact of the Qweak experiment will be: a
direct measurement of the proton's weak charge in a simple system; 
a theoretically clean method of measuring precisely the weak mixing angle
at low energy scale, a relatively unexplored regime of the Standard Model;
and a possible unambiguous indication of new physics.

\vspace{3\parskip}

\centerline{\bf THE QWEAK EXPERIMENT}

\begin{figure}[htbp]
\includegraphics[width=\textwidth]{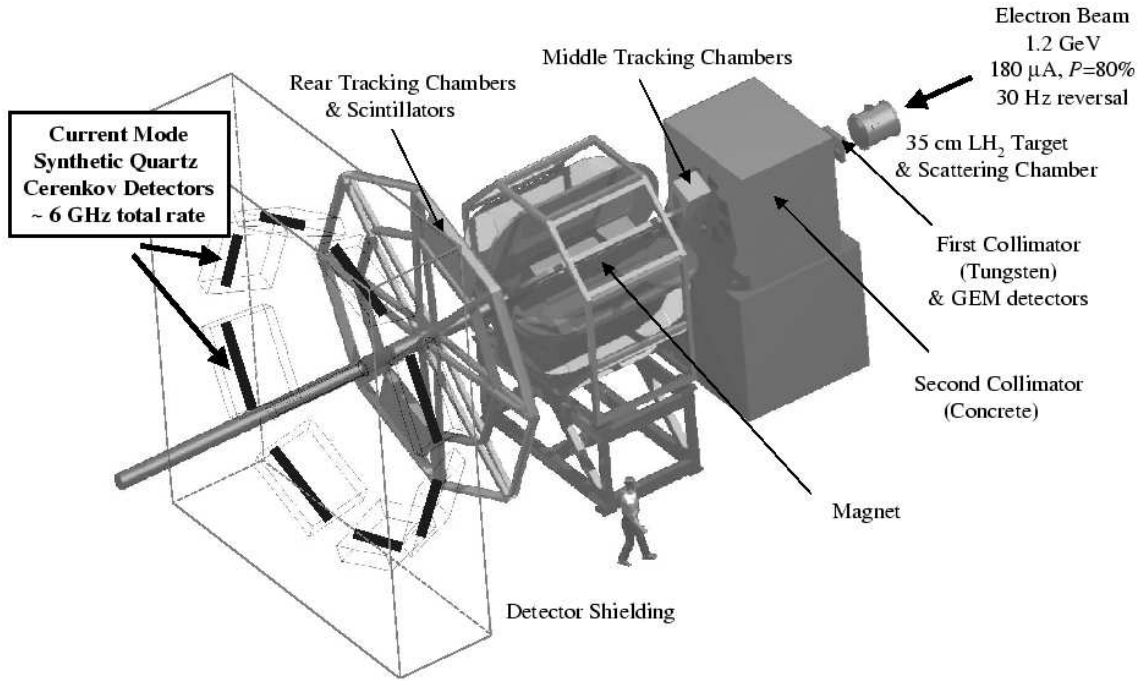}
\caption{Conceptual design for the Qweak experimental setup, in Hall C
at Jefferson Lab.  The eight quartz detectors are 
each 2 m$\times$12 cm$\times$2.5 cm. The spectrometer provides
clean separation of elastic and inelastic electrons at its focal plane.}
\label{fig:exp}
\end{figure}

An illustration of the conceptual design of the Qweak experiment is 
shown in Fig.~\ref{fig:exp}.  The experiment consists of 
scattering of a longitudinally polarized 1.2 GeV electron beam by
a 35~cm liquid hydrogen target.  Elastically scattered electrons 
at $9\pm2$ degrees are selected by a collimation system, and
the electrons are focused by a large toroidal magnet
onto a set of eight synthetic quartz {\u C}erenkov detectors.  
At the average experimental momentum transfer of
$Q^2 = 0.03\ {\rm GeV}^2$ the expected Qweak asymmetry 
is small, \mbox{-0.3} parts per million (ppm).
%To achieve this statistical precision given expected beam 
%polarization and detector performance, $3 \times 10^{16}$ 
%events will be acquired in 2200 hours of running time at JLab.
The expected event rate for scattered electrons of $\sim 6$ GHz 
(760~MHz per detector octant) precludes counting the individual
events.  Instead, the experiment will use current mode detection
and low noise front end electronics.
%Simulations indicate that the quartz will provide enough 
%{\u C}erenkov light per electron to have a significant
%number of photoelectrons produced ($\sim 50$ per event) 
%in the 130~mm PMT mounted on each end of each bar.

A tracking system will be used with the beam current reduced
by four orders of magnitude, allowing individual events to 
be observed.  This will enable both a measurement of the dilution of
the {\u C}erenkov detector signal by background and 
a precise determination of the average $Q^2$. 
The tracking system components will rotate to cover all octants, and will
include the following three sets of detectors:
a gas electron multiplier as a front vertex detector; wire chambers
near the magnet entrance as a measure of scattering angle;
and, at the focal plane, vertical drift chambers to allow mapping of
the analog response of the {\u C}erenkov system, and large scintillators 
as a charged particle trigger.

To extract the physics of interest for Qweak requires an extrapolation
to low $Q^2$ of hadronic form factors, $B(Q^2)$ above, which will be 
the dominant systematic error for the experiment at 2\%.
These form factors will contribute approximately one third of the 
experimental asymmetry.  They are known with reasonable precision at 
higher $Q^2$, and are suppressed by a factor of $Q^2$ with respect 
to the weak charge contribution for Qweak.  Other experiments
completed or underway at JLab (G$^0$, HAPPEX, HAPPEX II), 
Bates at MIT (SAMPLE), and MAMI at Mainz (A4), will be used to 
constrain these form factors.
%and their anticipated results should
%be sufficient for a 2\% uncertainty in the extrapolated 
%contribution to the Qweak measurement.  
If necessary, in the future the Qweak collaboration could pursue an 
independent measure of $B(Q^2)$ by running at a $Q^2$ different 
from 0.03~GeV$^2$.

Precision beam polarimetry is required in order to have a polarization
contribution to systematic error of less than 1.5\%, and a Compton 
polarimeter will be constructed in JLab Hall C.
Radiative corrections for the polarized $e-p$ scattering process at
these kinematics have been calculated and are known with 
a precision of 0.8\%~\cite{Erler}.

The Qweak physics proposal was approved by the January 2002 JLab 
PAC with an `A' scientific rating, and Qweak has become an 
important new thrust of the JLab scientific program.  The experiment 
presented a successful technical design review in January 2003.
The Qweak experiment will proceed in two stages: a statistics limited 
run with a low power target to achieve an 8\% or better result on the 
asymmetry in 2007, followed by a run of 2200 hours at 180~$\mu$A to 
achieve a 4\% result.  The absolute limits of the technique are under 
study.  This research was supported in part by the 
U.S.\ Department of Energy, the National Science Foundation,
and the Natural Sciences and Engineering Research Council of Canada.

%%%%%%%%%%%%%%%%%%%%%%%%%%%%%%%%%%%%%%%%%%%%%%%%
%% BACKMATTER
%%%%%%%%%%%%%%%%%%%%%%%%%%%%%%%%%%%%%%%%%%%%%%%%

\noindent
{\small
%%Qweak Collaboration List as of 18-Aug-2003
\begin{raggedright}

\noindent
$^\dagger$The Qweak Collaboration:\\
D.S.~Armstrong$^1$, 
T.D.~Averett$^1$, 
J.~Birchall$^2$, 
J.D.~Bowman$^3$, 
R.D.~Carlini$^4$, 
S.~Chattopadhyay$^4$, 
C.A.~Davis$^5$, 
J.~Doornbos$^5$, 
J.A.~Dunne$^6$, 
R.~Ent$^4$, 
J.~Erler$^7$, 
W.R.~Falk$^2$, 
J.M.~Finn$^1$, 
T.A.~Forest$^8$, 
D.J.~Gaskell$^4$, 
K.H.~Grimm$^1$, 
C.~Hagner$^9$, 
F.W.~Hersman$^{10}$, 
M.~Holtrop$^{10}$, 
K.~Johnston$^8$, 
R.T.~Jones$^{11}$, 
K.~Joo$^{11}$,  
C.~Keppel$^{12}$, 
E.~Korkmaz$^{13}$, 
S.~Kowalski$^{14}$, 
L.~Lee$^2$, 
A.~Lung$^4$, 
D.~Mack$^4$, 
S.~Majewski$^4$, 
G.S.~Mitchell$^3$, 
H.~Mkrtchyan$^{15}$, 
N.~Morgan$^9$, 
A.K.~Opper$^{16}$, 
S.A.~Page$^2$, 
S.I.~Penttila$^3$, 
M.~Pitt$^9$, 
M.~Poelker$^4$, 
T.~Porcelli$^{13}$, 
W.D.~Ramsay$^2$, 
M.J.~Ramsey-Musolf$^{17}$, 
J.~Roche$^4$, 
N.~Simicevic$^8$, 
G.R.~Smith$^4$, 
R.~Suleiman$^{14}$, 
S.~Taylor$^{14}$, 
W.T.H.~van Oers$^2$, 
S.B.~Wells$^8$, 
W.S.~Wilburn$^3$,
S.A.~Wood$^4$, 
C.~Zorn$^4$

\noindent
$^1${College of William and Mary, Williamsburg, Virginia 23187} \\
$^2${University of Manitoba, Winnipeg, MB, Canada R3T 2N2} \\
$^3${Los Alamos National Laboratory, Los Alamos, New Mexico 87545} \\
$^4${Thomas Jefferson National Accelerator Laboratory, Newport News,
Virginia  23606} \\
$^5${TRIUMF, Vancouver, BC, Canada  V6T 2A3} \\
$^6${Mississippi State University, Mississippi  39762} \\
$^7${Instituto de F\'\i sica, Universidad Nacional Aut\'onoma
      de M\'exico, 04510 M\'exico D.F., M\'exico} \\
$^8${Louisiana Technical University, Ruston, Louisiana  71272} \\
$^9${Virginia Polytechnic Institute and State University, Blacksburg,
Virginia  24061} \\
$^{10}${University of New Hampshire, Durham, New Hampshire  03824} \\
$^{11}${University of Connecticut, Storrs, Connecticut  06269} \\
$^{12}${Hampton University, Hampton, Virginia  23668} \\
$^{13}${University of Northern British Columbia, Prince George, BC,
Canada V2N 4Z9} \\
$^{14}${Massachusetts Institute of Technology, Cambridge, Massachusetts
02139} \\
$^{15}${Yerevan Physics Institute, 375036 Yerevan, Armenia} \\
$^{16}${Ohio University, Athens, Ohio 45701} \\
$^{17}${California Institute of Technology, Pasadena, California 91125}\\

\end{raggedright}

}

\end{document}